\documentstyle[epsfig,longtable]{aipproc}

\begin{document}
\title {ORCHESTRATION OF STARBIRTH ACTIVITY IN DISK GALAXIES:
N\lowercase{ew} P\lowercase{erspectives} \lowercase{from}
 U\lowercase{ltraviolet} I\lowercase{maging}}

\author{William H. Waller$^{\dagger*}$, Theodore P. Stecher$^{*}$, 
and the \\
Ultraviolet Imaging Telescope (UIT) Science Team$^{*\ddagger}$}

\address{$^{\dagger}$Hughes STX Corporation (waller@stars.gsfc.nasa.gov)}

\address{$^*$NASA Goddard Space Flight Center \\
Laboratory for Astronomy and Solar Physics (LASP) \\
Code 680, Greenbelt, MD 20771}

\address{$^{\ddagger}$
Website = http://fondue.gsfc.nasa.gov/UIT/UIT\_Homepage.html\thanks{
UIT research is funded through the Spacelab Office at NASA 
Headquarters under Project number 440-51.}}

\maketitle

\begin{abstract}
Ultraviolet imaging of nearby disk galaxies reveals the star-forming activity 
in these systems with unprecedented clarity.  UV images recently obtained with 
the Shuttle-borne Ultraviolet Imaging Telescope (UIT) reveal a remarkable 
variety of star-forming morphologies.  The respective roles of tides, waves, 
and resonances in orchestrating the observed patterns of starbirth activity are 
discussed in terms of the extant UV data.
\end{abstract}

\section* {WHY ULTRAVIOLET IMAGING?}

Despite the potential for obscuration by dust, a significant fraction 
of a disk galaxy's UV 
emission manages to escape and thus be detected by instruments located beyond 
the Earth's atmosphere.  Even in disk galaxies of high inclination, UV imaging 
can reveal widespread emission 
(cf. Fanelli et al., Neff et al., \& O'Connell, 
these Proceedings).

Ultraviolet imaging of nearby disk galaxies reveals the star-forming activity 
with unprecedented clarity.  
Unlike imaging at H$\alpha$, UV 
imaging {\it directly} traces the full range of OBA stellar spectral types, 
thereby sampling 
the recent-epoch ``Population I'' component of each galaxy more completely.
UV imaging also provides a cleaner separation of the hot star component 
in regions dominated by cooler stars (e.g. central disks \& bulges).  Moreover, 
UV imaging is unaffected by line absorption in the atmospheres of B \& A-type 
stars --- unlike imaging at H$\alpha$ and H$\beta$ --- thereby providing a 
truer representation, where these populations are concentrated (see FIGURE 1).
Finally, the 
UV 
colors of OB/HII regions can be used to derive extinction-free UV 
luminosities \cite{Hill95}\cite{Hill96}, 
whereas few constraints exist for deriving 
extinction-free EUV (e.g. Lyman continuum) luminosities 
based on H-line, radio-continuum, or other 
indirect 
indices.  

\begin{figure} 
\centerline{\epsfig{file=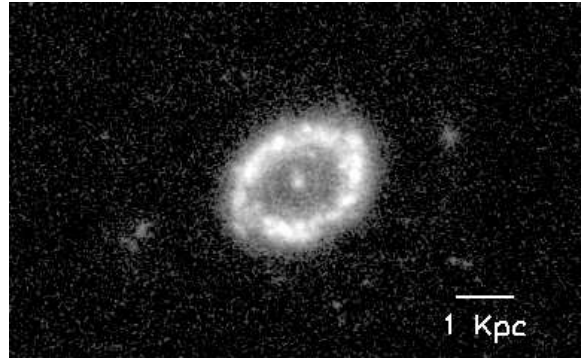}}
\caption{FUV ($\lambda$1520) image of the Sab galaxy M94 (inner disk), showing 
bisymmetric star-forming knots, a resonant ring of starburst activity and 
diffuse FUV emission interior to the ring --- where H$\alpha$ is {\it in 
absorption} due to the underlying B \& A-type stellar populations.}
\end{figure}

\section* {UV MORPHOLOGIES OF DISK GALAXIES}

UV images of nearby disk galaxies obtained with the 
Shuttle-borne {\it Ultraviolet Imaging Telescope} (UIT) 
reveal a remarkable variety 
of star-forming morphologies.
These Pop I patterns yield important insights 
to the respective roles of {\it tides, waves, and resonances} in orchestrating 
starbirth activity in disk galaxies.

A comparison of M33 (Scd), M74 (Sc), and M81 (Sb) 
at UV and visible wavelengths 
highlights the Pop I character of the UV imagery 
\cite{Waller94}.
{\it Flatter radial distributions} are evident in the UV --- with 
exponential 
scalelengths that are 20--45\% larger.
The effects of reddening, abundance, and IMF variations do not fully explain 
the differences.  
The 
flatter UV profiles most likely 
indicate that the median radius of 
star-forming activity has migrated outward over the past several Gyrs 
\cite{Cornett94}.  

The UV morphologies also show {\it narrower arms} --- delineated by 
a combination of direct starlight from 
OBA associations and indirect (scattered) radiation from dust 
associated with the massive young stars \cite{Hill95}\cite{Waller96}.
The narrower UV features indicate that 
star formation over the past 
$\sim$10 Myrs occupies a significantly smaller areal domain 
than the $\sim$1--1000 Myr legacy of star formation that is traced at longer 
wavelengths.

\subsection* {Tides}

In the giant ScI spiral M101, 
multiple linear arm segments ({\it ``crooked 
arms''}) can be traced throughout the disk (see FIGURE 2).  
These features, along with a faint 
spiral arm and {\it ``curly tail''}
feature that links the outermost supergiant 
HII region with the rest of the galaxy, indicate that {\it tidal processes of 
both external and internal origin} are directing the current starbirth 
activity \cite{Waller96}.

\begin{figure} 
\centerline{\epsfig{file=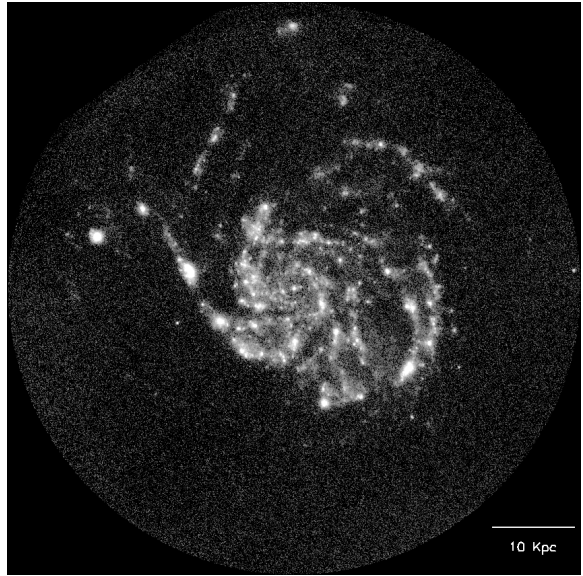}}
\caption{FUV image of the giant ScI galaxy M101, showing multiple ``crooked 
arms'' throughout the disk and a faint outer arm terminating in 
a ``curly tail'' feature.}
\end{figure}

Numerical simulations of isolated disk galaxies show that the 
``crooked arm'' behavior can arise through the action of ``massive disturbers'' 
orbiting within the disks.  The outermost supergiant HII region, NGC 5471 may 
represent one of these massive disturbers.  Larger-scale morphological and 
kinematic anomalies in M101, 
including the faint arm and ``curly tail'' feature, 
require external interactions with companion 
galaxies.  Such interactions can induce the formation of massive condensations 
at the ends of the tidal tails, perhaps explaining the origin of NGC 5471 at 
the terminus of M101's ``curly tail''
\cite{Waller96}.  Similar behavior can be found in 
other giant Sc galaxies with ``companions,'' including M51, NGC 1232, NGC2805, 
and NGC 4303.

\subsection* {Waves}

In the ``grand-design'' ScI galaxy M74 (NGC 628), reflection of the 
UV-emitting disk upon itself shows the spiral structure to be more symmetric 
than is observed at visible wavelengths --- thus arguing for large-scale 
dynamics (e.g. density waves) governing the current-epoch star formation 
\cite{Chen92}. 

Evidence for {\it spatio-temporal sequences} of molecular-cloud 
aggregation, massive star 
formation, cluster evolution, 
and cloud disruption can be found in M74, M51 (O'Connell, these Proceedings), 
and the {\it inner disk} of M101 \cite{Waller96}.
In M101, far-UV emission is often found on the 
outer (downstream) side of the inner-disk CO arms.  Modeling the FUV--CO 
displacements according to density-wave dynamics results in a wave pattern 
speed and co-rotation radius remarkably similar to those derived from a 
multi-mode analysis of the optical spiral structure 
\cite{Waller96}.  
Similar downstream displacements between the FUV and H$\alpha$ emission is 
evident in the 
SE arm of M74, again indicating density-wave dynamics at work.

\subsection* {Resonances}

In the Sab spiral M94 (NGC 4736), UV imaging reveals an inner starbursting 
ring and bi-symmetric outer knots in high contrast against the underlying 
visible bulge and disk (see FIGURE 1).  
Dynamical resonances seem to best explain these 
transient features.  Similar UV rings are evident in the inner disks 
of NGC 1317 (SBa), NGC 1512 (SBb), NGC 3351 (SBb), and M100 
(NGC 4321) 
(Sc) --- most of which are also of ``early'' morphological type.
  
Resonances may also explain the dearth of UV emission 
interior to M81 and M31's ring-like 
spiral arms \cite{Hill92}, 
whereby star-forming gas migrates outwards from the interior and 
piles up near the Inner Lindblad 
Resonances (Kenney, these Proceedings).
Because the longer-wavelength emission traces the older stellar 
populations and is so prominent interior to the 
ringlike arms, we conclude that the resonant locations and 
their constructive/inhibiting effects must have evolved with time.

\end{document}